\documentclass[twocolumn,aps,prr,%showpacs,
superscriptaddress
]{revtex4}

\usepackage{graphicx,color,hyperref}

\usepackage{amsmath}
\usepackage{amsfonts}
\usepackage{amssymb}
\usepackage{dsfont}

\DeclareMathOperator\erfc{erfc}

\usepackage{braket}

\newcommand{\tr}{\mathrm{tr}}

\usepackage[normalem]{ulem} % can be commented out unless we use \sout

\usepackage[normalem]{ulem} % can be commented out unless we use \sout

\usepackage{standalone}
\usepackage{tikz}
\usetikzlibrary{positioning}

\def \be {\begin{equation}} 
\def \ee {\end{equation}} 

\def \bea {\begin{eqnarray}} 
\def \eea {\end{eqnarray}}

\def \9 {\Bigg(}
\def \0 {\Bigg)}

\begin{document}

\title{Subsystem Trace-Distances of Two Random States}

\author{ 
Joaquim 
Telles de Miranda 
}
\affiliation{Centro Brasileiro de Pesquisas F\'isicas, Rua Xavier Sigaud 150, 22290-180, Rio de Janeiro, Brazil 
} 

\author{ 
Tobias Micklitz  
}
\affiliation{Centro Brasileiro de Pesquisas F\'isicas, Rua Xavier Sigaud 150, 22290-180, Rio de Janeiro, Brazil 
} 

\date{\today}

\begin{abstract}
We study two-state discrimination in chaotic quantum systems. 
Assuming that one of two $N$-qubit pure states 
has been randomly selected, the probability to correctly identify the selected state
from an optimally chosen experiment involving a subset of $N-N_B$ qubits is 
given by the trace-distance of the states, with $N_B$ qubits partially traced out. 
In the thermodynamic limit $N\to\infty$, 
the average subsystem trace-distance 
for random pure states makes a sharp, first order 
transition from unity to zero at $f=1/2$, 
as the fraction $f=N_B/N$ of unmeasured qubits is increased.
We analytically calculate the  corresponding crossover 
for finite numbers $N$ of qubits,  
study how it is affected by the 
presence of local conservation laws, 
 and test  our predictions against exact diagonalization of 
models for many-body chaos. 
\end{abstract}

\maketitle

\section{ Introduction} 

 The capability for storing and processing quantum information relies  
 on the ability to discriminate between quantum states.   
 Quantifying distinguishability of quantum states, 
 specifically when only access to subregions is available, 
 is thus a problem of fundamental and practical interest. 
In chaotic systems initially localized information is `scrambled' 
into the many degrees of freedom of the 
system,  and subsystem density matrices of generic pure states are nearly indistinguishable from fully thermal states. 
The `nearly' was specified by Page for pure random states, which also serve as proxies for  
eigenstates of chaotic systems,  
in his classic paper~\cite{Page1993}.  
There he showed that for these states the information stored in the smaller subsystem, defined as the 
deficit of the subsystem entanglement entropy $S_A$ from the maximum entropy 
of a fully mixed state, 
is on average 
$I_A\equiv \ln D_A-S_A = D^2_A/2D$.
Here $D_A$, and $D$ are the Hilbert space dimensions of the smaller subsystem, 
respectively, the joint system, and corrections small in $1/D$ have been neglected. 
For reasonably large systems the entanglement entropy is self averaging
and the average $S_A$ is also typical~\cite{Bianchi2019,RigolKieburg2021}. 
Conservation laws generally reduce entanglement, increasing thus the capability to store 
information~\cite{Bianchi2019,Rigol2008,Lau2020,Ares2022,VidmarRigol17,Sugiura18,Sugiura12,Sugiura13,Calabrese2022}. 
If multiple charges are conserved, entanglement is (on average) promoted if the charges fail to commute with each other. That is, 
 Page curves for non-commuting charges lie above that of commuting charges, 
 as recently pointed out in Ref.~\cite{Halpern2022}.  
Morevover, in the presence of locally conserved charges, 
the entire charge distributions of initial states are conserved. 
The largest amount of information can then 
be stored in states with the broadest charge distribution~\cite{AltlandHuseMicklitz2022}.

While Page's formula provides information on subsystems, 
it does not give an answer to state-discrimination in  fully information scrambling systems. 
Specifically, imagine one of two known random pure states $\rho$, $\sigma$, both 
composed of $N$ qubits, has been 
randomly selected. What is the (average) probability $P_{\rho\sigma}$ that performing an optimally 
chosen experiment on $N_A$ of the qubits we correctly identify the selected 
state, and how is $P_{\rho\sigma}$ affected by conservation laws? 
In this paper we want to investigate these questions, and 
the outline  is as follows: We start briefly reviewing in Section~\ref{sec1} the concepts of
the trace-distance $D_1$, its generalization the Schatten $n$-distances $D_n$, Page states, 
and the calculation of  
$D_1$ from $D_n$ via the replica trick. 
We then discuss in Section~\ref{sec2} 
the combinatorics involved in the calculation of 
average subsystem Schatten $n$-distances of random pure states. 
In Section~\ref{sec3} we analyze the average subsystem trace-distances 
of random pure states %, 
and consequences of local conservation laws. 
We conclude in Section~\ref{sec5} with a summary and discussion,  
and give further technical details in the Appendices.

\section{ Schatten-distances, Page states, and replica trick}
\label{sec1}

Optimal quantum state discrimination is generally challenging,
and the only completely analyzed case is for two states, see e.g. Ref.~\cite{Bae2015} for a review.  
The trace-distance provides a natural metric for two-state discrimination. 
According to the Holevo–Helstrom theorem the best success 
probability for the latter is encoded in the 
 $1$-Schatten- or trace-distance as 
$P_{\rho\sigma}=\frac{1}{2}\left(1 +D_1(\rho,\sigma)\right)$~\cite{footnote1}.  
General Schatten $n$-distances here are defined as
$D_n(\rho,\sigma)\equiv \frac{1}{2^{1/n}}||\rho - \sigma||_n$,  
with $n$-norm 
of a matrix $\Lambda$ determined by its singular values $\lambda_i$ as
$||\Lambda||_n= \left(\sum_i \lambda^n_i\right)^{1/n}$~\cite{footnote2}.
Notice that   
all Schatten distances are symmetric in the inputs, 
positive semi-definite, equal to zero if and only if inputs are identical, 
and obey the triangular inequality. That is, they all satisfy the properties of a metric, 
with normalization here chosen such that $0\leq D_n \leq 1$.  
Our focus here is on two-state discrimination and we thus concentrate on the $1$-distance.

Consider then a $D$-dimensional Hilbert space with 
entanglement-cut bi-partitioning the total system into subsystems $A$, $B$ 
of dimensions 
$D_A$ and $D_B$, respectively, with $D_AD_B=D$. 
Without much loss of generality, we focus here on  
$N$ qubit systems parametrized by
 $(a,b)$, where the $N_A$-bit vector $a$ labels the $D_A = 2^{N_A}$ states of subsystem $A$ and $b$ 
the $D_B = 2^{N_B}$ states of $B$, with $N_B=N-N_A$. 
 Following Page, we then consider two random pure states 
$|\psi^\alpha\rangle = \sum_{a, b} \psi^\alpha_{ab}|a, b\rangle$, 
$\alpha=\rho,\sigma$, 
with 
Gaussian distributed complex amplitudes $\psi^\alpha_{ab}$, 
chosen to have zero mean and variances
\begin{align}
\label{eq:variances}
\langle\psi_{ab}^{\alpha}\bar{\psi}_{cd}^{\beta}\rangle 
&=
\frac{1}{D}\delta_{ac}\delta_{bd}\delta_{\alpha\beta}. 
\end{align}
$|\psi^\alpha\rangle$ describe infinite temperature thermal states of generic chaotic systems, 
and using Eq.~\eqref{eq:variances} we employ that 
correlations induced by the normalization constraint are negligible 
for reasonable large systems $D\gg1$. 
Tracing out subsystem $B$, information is lost
and mixedness of the reduced density matrices, 
\begin{align}
\label{eq:reduced_density_matrices}
\rho_A&={\rm tr}_B(|\psi^\rho\rangle\langle\psi^\rho|),
\quad
\sigma_A={\rm tr}_B(|\psi^\sigma\rangle\langle\psi^\sigma|),
\end{align}
increases with the number of partially traced qubits $N_B$. 

\begin{figure}[t!]
\centering
\includegraphics[width=.45\textwidth]{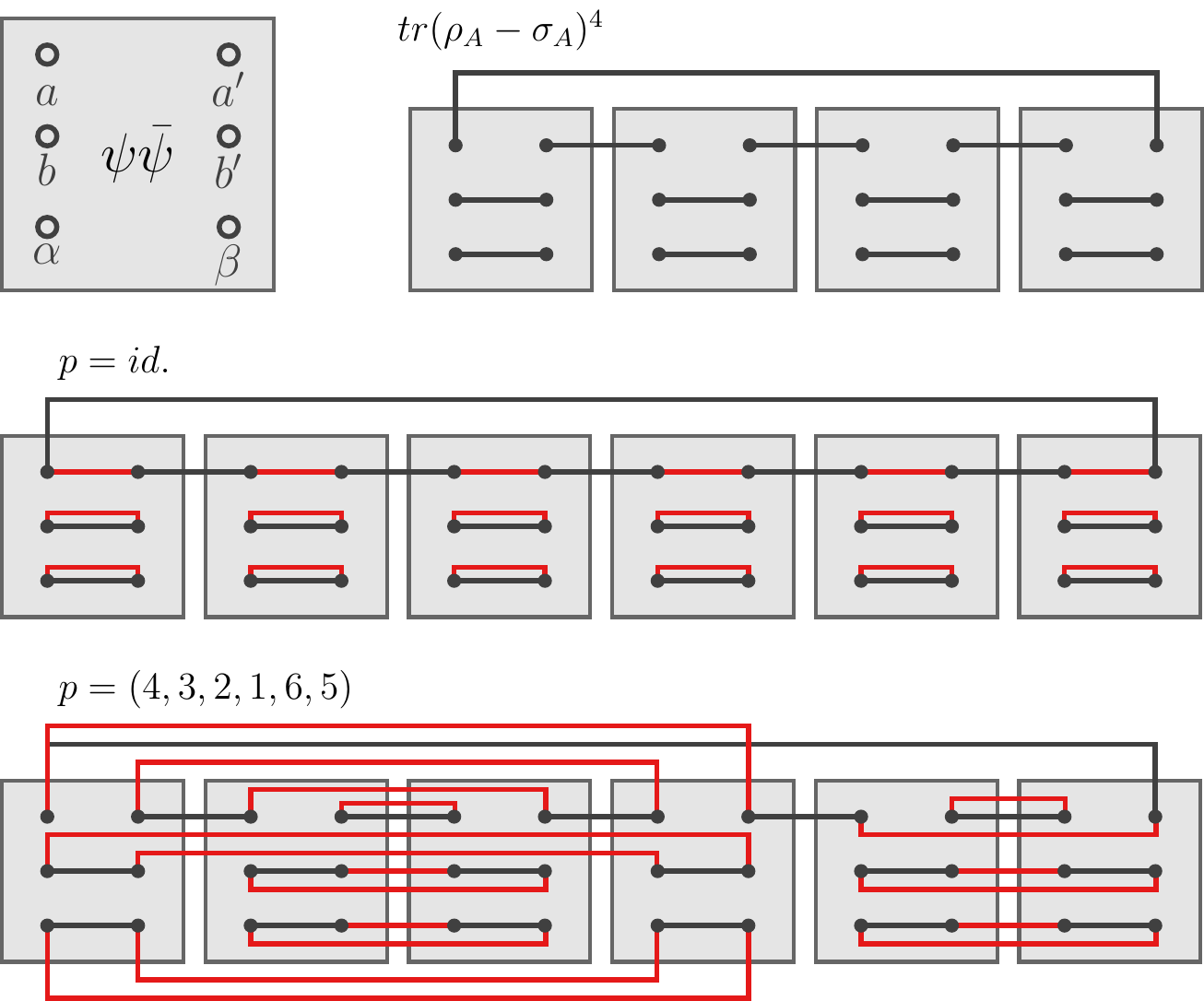}
\caption{
 Tensor network representation of averages Eq.~\eqref{eq:Schatten_distances_wf}. 
Top left: Representation of a pair of amplitudes from the expansion
$\alpha_A = \sum_{a, a'} \sum_b \psi^\alpha_{ab} \bar{\psi}^\alpha_{a'b}|a\rangle\langle a'|$, 
with $\alpha=\rho, \sigma$. Each dot represents an index to be contracted, 
and contractions must be between right- and left-side indices.
Top right: Structure of $\tr(\rho_A-\sigma_A)^4$, with black lines representing index contractions 
resulting from matrix multiplication in subspace $A$ (top line), 
traces in subspace $B$ (middle line), and state indices $\alpha=\rho,\sigma$ (bottom line). 
Notice that the index structure of states  follows that of 
subspace $B$. 
 When averaging the trace we only consider 
leading order  contributions in $1/D$ 
resulting from non-crossing permutations. Gaussian averages then 
 impose further contractions upon the structural ones, indicated by the red lines.
Middle: Resulting index structure for $n=6$ and identity permuation $p=id$. 
This establishes six $B$-cycles each consisting of one element, 
i.e. in the notation of main text 
$\Lambda_6=(1^6,2^0,3^0,4^0,5^0,6^0)$. 
Bottom: Another example of a non-crossing 
permutation for $n=6$, $p=(4, 3, 2, 1, 6, 5)$. 
This permutation establishes three $B$-cycles  each consisting of two elements, 
$\Lambda_6=(1^0,2^3,3^0,4^0,5^0,6^0)$. 
In the middle diagram, contributions from states
$\rho$ (positive sign) and $\sigma$ (negative sign) sum to zero
 in each of the one-element cycles. 
In the bottom diagram contributions from 
$\rho$ and $\sigma$ both come with positive sign and sum to two, i.e. contributions from the
three cycles add up to $2^3=8$.
}
\label{fig1}
\end{figure}

To find trace-distances of Eq.~\eqref{eq:reduced_density_matrices}
we employ the replica trick recently discussed in Ref.~\cite{Calabrese2019}.  
We first calculate Schatten-distances for general even integer $n$, 
analytically continue to real $n$, 
and finally 
take the limit $n$ 
to unity, 
\begin{align}
\label{eq:trace_distance_replica_trick}
\langle 
D_1(\rho_A,\sigma_A)
\rangle
&=
\frac{1}{2}\lim_{n\to1} 
\langle
{\rm tr}\left(
\rho_A-\sigma_A\right)^n 
\rangle.
\end{align}
Restricting to even integers here is important, since corresponding expression for odd integers vanish 
(see also below), and a replica limit for the latter is thus trivially zero~\cite{Calabrese2019}. 
Expanding powers in Eq.~\eqref{eq:trace_distance_replica_trick}, 
we are confronted with the averages 
\begin{align}
\label{eq:Schatten_distances_wf}
&
\langle
{\rm tr}_A(\rho_A-\sigma_A)^{n}
\rangle 
\nonumber\\
&\,\,
= 
{\rm sgn}(\sigma)
\langle
\psi_{a^1b^1}^{\alpha_1}\bar{\psi}_{a^2b^1}^{\alpha_1}
\psi_{a^2b^2}^{\alpha_2}\bar{\psi}_{a^3b^2}^{\alpha_2}
\cdot\cdot\cdot
\psi_{a^{n}b^{n}}^{\alpha_{n}}\bar{\psi}_{a^1b^{n}}^{\alpha_{n}}
\rangle,
\end{align}
where sums over repeated indices $\alpha_i=\rho,\sigma$, 
$a_i=1,..,D_A$, $b_i=1,...,D_B$ are implicit, and  
the sign-factor is ${\rm sgn}(\sigma)=\pm 1$ if an 
even/odd number of density matrices $\sigma_A$ 
is involved in the product. 
Following previous works~\cite{MonteiroPRL2021}  
the bookkeeping of index configurations entering the products
is conveniently done in a tensor network representation shown in Fig.~\ref{fig1}. 
The solid lines here indicate how the indices of matrices 
$\left(\psi \bar \psi\right)^{\alpha\beta}_{ab;a'b'}=\psi^\alpha_{ab} \bar \psi^\beta_{a'b'}$
 are constrained due to matrix multiplication in subspace $A$,  subsystem-traces over $B$,  
and state indices $\alpha= \rho,\sigma$, respectively  (see also figure caption). 
Further constraints then arise from %the 
Gaussian averages Eq.~\eqref{eq:variances}.
These are indicated by the red lines, 
keeping track of Hilbert space and state indices after contractions. 
For Page states, each of the $n!$ contributions resulting from the Gaussian averages 
of the $2n$ complex amplitudes in Eq.~\eqref{eq:Schatten_distances_wf}
is weighted by an overall factor $1/D^{n}$,  
and terms can be organized according to the numbers of free subspace summations, or `cycles', 
as we discuss next.

%%%%%%%%%%%%%%%%%%%%%%%%%%%%%%%%%%%%%%%%%%%%%%%%%%%%%%%%
%%%%%%%%%%%%%%%%%%%%%%%%%%%%%%%%%%%%%%%%%%%%%%%%%%%%%%%%

\section{ Combinatorics of averages}
\label{sec2}

It is instructive to  
first focus on the contribution involving only a single state 
density matrix, say 
$\rho_A$, 
\begin{align}
\label{eq:moment_one_rdm}
\langle
\mathrm{tr}_A(\rho_A^{n})
\rangle
&=
\langle
\psi^\rho_{a^1 b^1}\bar \psi^\rho_{a^2 b^1}
\psi^\rho_{a^2 b^2} \dots \psi^\rho_{a^{n}b^{n} }\bar \psi^\rho_{a^1 b^{n}}
\rangle.
\end{align}
These averages have 
been recently discussed in the context of the entanglement entropy~\cite{footnote3},
and the main observations are~\cite{Stanford2019,Liu20}:
(i) the $n!$ contributions resulting from the average of $2n$ Gaussian distributed complex variables 
can be organized as a sum over the permutation group
$\langle
\mathrm{tr}_A(\rho_A^{n})
\rangle
=
\frac{1}{D^n}\sum_{p\in S_n} D_A^{C(\pi^{-1}\circ p)}D_B^{C(p)}$, 
where  $C(p)$ is the number of cycles in the permutation $p$ and $\pi$ defined by 
$\pi(i)=(i+1){\rm mod}(n)$, 
(ii) the maximal number of cycles is $C(\pi^{-1}\circ p)+C(p)=n+1$ 
and is realized by the non-crossing permutations, (iii) 
their combinatorics is encoded in the Narayana numbers $N(n,k)$ 
where $k$ the number of cycles in $B$, and (iv) contributions of crossing permutations  
are suppressed in powers of $1/D$. 
Neglecting the latter, one thus arrives at
$\langle
\mathrm{tr}_A(\rho_A^{n})
\rangle
= \frac{1}{D^{n}}
\sum_{k=1}^{n}
N(n,k)
D_A^{n-k+1} D_B^k$, which can be further organized in power-series 
defining hypergeometric functions. 
The calculation of Eq.~\eqref{eq:moment_one_rdm} has thus been 
succeeded once the Narayana numbers have been identified as 
the combinatorial coefficients 
 summing all possible non-crossing permutations of the  $n$ elements 
 containing $k$ $B$-cycles. 

To extend the calculation to all terms Eq.~\eqref{eq:Schatten_distances_wf} 
we introduce the Kreweras numbers,
\begin{align}
\label{eq:Kreweras}
{\rm Krew}(\Lambda_n)
&=
\frac{n!}{\lambda_1!...\lambda_n!(n+1-l(\Lambda_n))!},
\end{align}
with $l(\Lambda_n)\equiv\sum_{i=1}^{n} \lambda_i$.
They count the number of non-crossing permutations 
composed of $\lambda_i$ $B$-cycles formed of $i$ elements, 
$\Lambda_n=(1^{\lambda_1},2^{\lambda_2},...,n^{\lambda_n})$, 
and thus provide more detailed information than the Narayana numbers \cite{KREWERAS1972333, SIMION2000367}. 
Specifically, Narayana numbers sum all Kreweras numbers 
specified by $l(\Lambda_n)=k$ and $\sum_{i=1}^{n} i\lambda_i =n$ (see also Appendix~\ref{app_Kreweras}). 
With this additional information we 
are now ready to tackle the combinatorics required for the calculation 
of average Schatten $n$-distances.

Cycles in Eq.~\eqref{eq:Schatten_distances_wf}
can be formed from states $\rho$ or $\sigma$.
Both contribute the same in absolute value, however, 
not always with same sign. That is, while 
the sign is always positive for $\rho$-cycles 
it alternates for $\sigma$-cycles
depending on whether the number of elements involved in the cycle
is even/odd~\cite{footnote5}. 
For permutations with odd element cycles the contributions 
from $\rho$-cycles 
and $\sigma$-cycles  
thus cancel, and therefore only 
those consisting of even elements contribute. 
Noting that $b$-indices follow that of state-indices 
(see also Fig.~\ref{fig1}) we arrive at the same conclusion for $B$-cycles.
 That is, for non-crossing permutations with $k$ $B$-cycles  
 only $\Lambda_n=(1^0,2^{\lambda_2},3^0...,n^{\lambda_n})$ 
  with all cycles composed of even elements contribute, 
  and summing the two choices 
 $\alpha=\rho,\sigma$ for each of the $k$ cycles adds up to a factor $2^k$. 
As a corollary we notice that moments in Eq.~\eqref{eq:Schatten_distances_wf} 
involving odd powers $n$ vanish, as anticipated above. 
We are then left with the combinatorical task of counting 
the number of $k$-cycles consisting only of even elements. 
Summing the corresponding Kreweras numbers (see Appendix~\ref{app_Kreweras} for details), 
we find 
\begin{align}
N_e(n,k) 
&= 
\frac{2}{n}\binom{n/2}{k}\binom{n}{k-1}.
\end{align}

\section{ Average trace-distances}
\label{sec3}

Joining parts, we find the average Schatten $n$-distances of two random Page states 
$\langle D_{n}(\rho_A,\sigma_A) \rangle
= 
\frac{1}{2D^{n}}
\sum_{k=1}^{n/2} 2^kN_e(n,k)D_A^{n-k+1}D_B^{k}$, 
which   
 can be organized into a hypergeometric function (see Appendix~\ref{app_subsystem_trace_distance} 
for details). For the latter the replica limit can be taken, 
and we arrive at   
the average trace-distance of two random Page states,  
\begin{align}
\label{eq:Main_Result}
\langle 
D_1(\rho_A, \sigma_A) 
\rangle
&= 
\begin{cases}
1-\frac{1}{4x}, 
&\quad x\geq 1,
\\
\frac{8\sqrt{x}}{3\pi}\,
{\cal F}(x), 
&\quad x \leq 1,
\end{cases}
\end{align}
where $x=\frac{D_A}{2D_B}$,  
 and
${\cal F}(x)={}_2F_1( \frac{1}{2},-\frac{1}{2},\frac{5}{2},x)$ 
a hypergeometric function. 
Similar results have been recently derived in Ref.~\cite{Ryu2021} using free probability techniques, and 
 Eq.~\eqref{eq:Main_Result} agrees with their asymptotic expressions.

Fig.~\ref{fig2} shows the trace-distance of Page states 
for different numbers of qubits $N$ (solid lines)
as a function of the fraction  $f\equiv N_B/N$ 
of partially traced qubits, and $x=\frac{1}{2}D^{1-2f}$. 
Increasing $N$, one observes a sharp transition from 
$\langle D_1\rangle\simeq 1$ 
to $\langle D_1\rangle\simeq 0$
once the partially traced system contains more than half of 
all qubits, $f>1/2$, 
where `$\simeq$' indicates 
equality up to corrections ${\cal O}(1/D)$ exponentially small 
in the number of qubits. 
In the thermodynamic limit, this becomes a first order transition, 
describing the emergence of self-averaging of the reduced density 
matrix of random pure states once more than half of the qubits 
are partially traced out. 
Two-state discrimination 
of Page states for $N\to \infty$ is therefore possible with unit probability 
$P_{\rho\sigma}=1$ if the  
measurement is performed on more than half of 
the qubits, but becomes essentially impossible, $P_{\rho\sigma}=1/2$,  
for measurements involving fractions smaller than half of the qubits.  
Finite $N$ curves all intersect in $f=1/2$, and the probability for two-state discrimination 
using measurements on half of the qubits is $P_{\rho\sigma}=\frac{5\pi+4}{8\pi}\simeq 0.78$, 
in agreement with previous work Ref.~\cite{Zyczkowski2016}.  
Expanding the trace-distance around 
$f=1/2$ one finds    
$\langle D_1\rangle\simeq \frac{4+\pi}{4\pi} + \frac{1}{4}(D^{1-2f}-1)$, i.e.  
the probability of two-state discrimination 
 decays with the number of unmeasured qubits as 
$P_{\rho\sigma} ={\rm const.} -N\ln2/4 \times (1-2f)$. 
At $f=1$ the trace-distance is trivially zero for any realization of states, while 
Eq.~\eqref{eq:Main_Result} predicts 
a value ${\cal O}(1/\sqrt{D})$, of the same order as the value at $f=1-1/N$. 
The erroneous finite value at $f=1$ reflects that  
 Page states  
Eq.~\eqref{eq:variances}  
 realize state-normalization only on average, rather than for each realization. 
We next turn to a discussion on the consequences of local conservation laws.

\subsection*{ Conservation laws}
\label{sec4}

Page’s random pure states describe generic quantum states in chaotic systems lacking any structure e.g. induced by
 conservation laws. For these information scrambling  is most efficient, and 
 we next discuss how local conservation of scalar charges, e.g. particle number, uni-axial magnetization, etc.,   
affects two-state discrimination in chaotic systems. 
More specifically, we consider the presence of a single, extensive conserved scalar operator $\hat Q$  
 that is subsystem additive: 
A partition of the system into the two subsystems $A$ and $B$ 
implies a decomposition $\hat{Q}=\hat{Q}_A+\hat{Q}_B$, and 
eigenstates $\hat Q |n\rangle= Q(n) |n\rangle$ 
can be labeled by 
$n=(a,b)$ 
with $Q(n)=Q_A(a)+Q_B(b)$. 
It is then convenient to introduce the spectral distribution 
 of $\hat Q$,  $F(Q)\equiv D\Omega(Q)=\sum_n\delta_{Q,Q(n)}$,
and corresponding subsystem spectral densities, $F_S(Q_S)\equiv D_S\Omega_S(Q_S)$, with $S=A,B$.
For most cases we can assume that (except from the far tails of the spectrum irrelevant for 
our considerations) the unit normalized spectral densities 
are well approximated by Gaussians,
\begin{align}
\label{eq:gaussian_spectral_density}
\Omega(Q)
&=
\Omega(Q,N)
=
\frac{1}{\sqrt{2\pi\gamma^2N}}
e^{-\frac{Q^2}{2\gamma^2 N}},
\end{align} 
with $\gamma$ some $N$-independent scale, 
and correspondingly for subsystem densities
$\Omega_S(Q_S)=\Omega(Q_S,N_S)$. 
For convenience we here choose $Q=0$ as the value 
with largest spectral weight. 

Focusing then on the two-state discrimination of charge eigenstates, 
we substitute the average for Page states, Eq.~\eqref{eq:variances},
for the symmetry refined version,
\begin{align}
\label{eq:variances_conservation_laws}
\langle
\psi^\alpha_{ab}\bar{\psi}^\beta_{cd}
\rangle
&=
\frac{1}{F(Q)}
\delta_{Q_A(a)+Q_B(b),Q}
\delta_{ac}\delta_{bd}\delta_{\alpha\beta}.
\end{align} 
Eq.~\eqref{eq:variances_conservation_laws} 
describes random wave functions 
in the subspace of dimension $F(Q)$ fixed by total charge $Q$, 
and $\delta_{Q_A(a)+Q_B(b),Q}$ introduces non trivial correlations between 
the subsystems. 
We first concentrate on the value with largest spectral weight
$Q=0$ for which the 
most universal behaviour can be expected. 
Following then the previous calculation for Page states,
we find (see Appendix~\ref{app_charge_eigenstates} for details)
\begin{align}
\label{eq:Q_eigenstate}
\langle 
D_1(\rho_A, \sigma_A) 
\rangle
&= 
\begin{cases}
1-\frac{1}{4x\sqrt{2f}}, 
&x\geq 1,
\\
\frac{8\sqrt{x_f}}{3\pi}
{\cal G}(x_f,f), 
&x \leq 1,
\end{cases}
\end{align}
where  we introduced 
$x_f \equiv  
x\sqrt{f/(1 - f)}$ with 
$x=\frac{D_A}{2D_B}$  
as before,  
and
${\cal G}(x,f)
=
\frac{3\pi}{4}\sum_{k=0}^\infty
c_k x^k
\binom{1/2}{k}\binom{1}{k+3/2}$ 
with 
$c_k=\left((1+2k)f+1/2-k\right)^{-1/2}$. 
As anticipated, the result Eq.~\eqref{eq:Q_eigenstate}  
is unviversal 
in the sense that it only depends on the numbers of 
qubits $N$, $N_B$.

%%%%%%%%%%%%%%%%%%%%%%%%%%%%%%%%%%%%%%%%%%%%%%%%%%%%%%%%
\begin{figure}[t!]
\centering
\vspace{-.5cm}
\includegraphics[width=8.7cm]{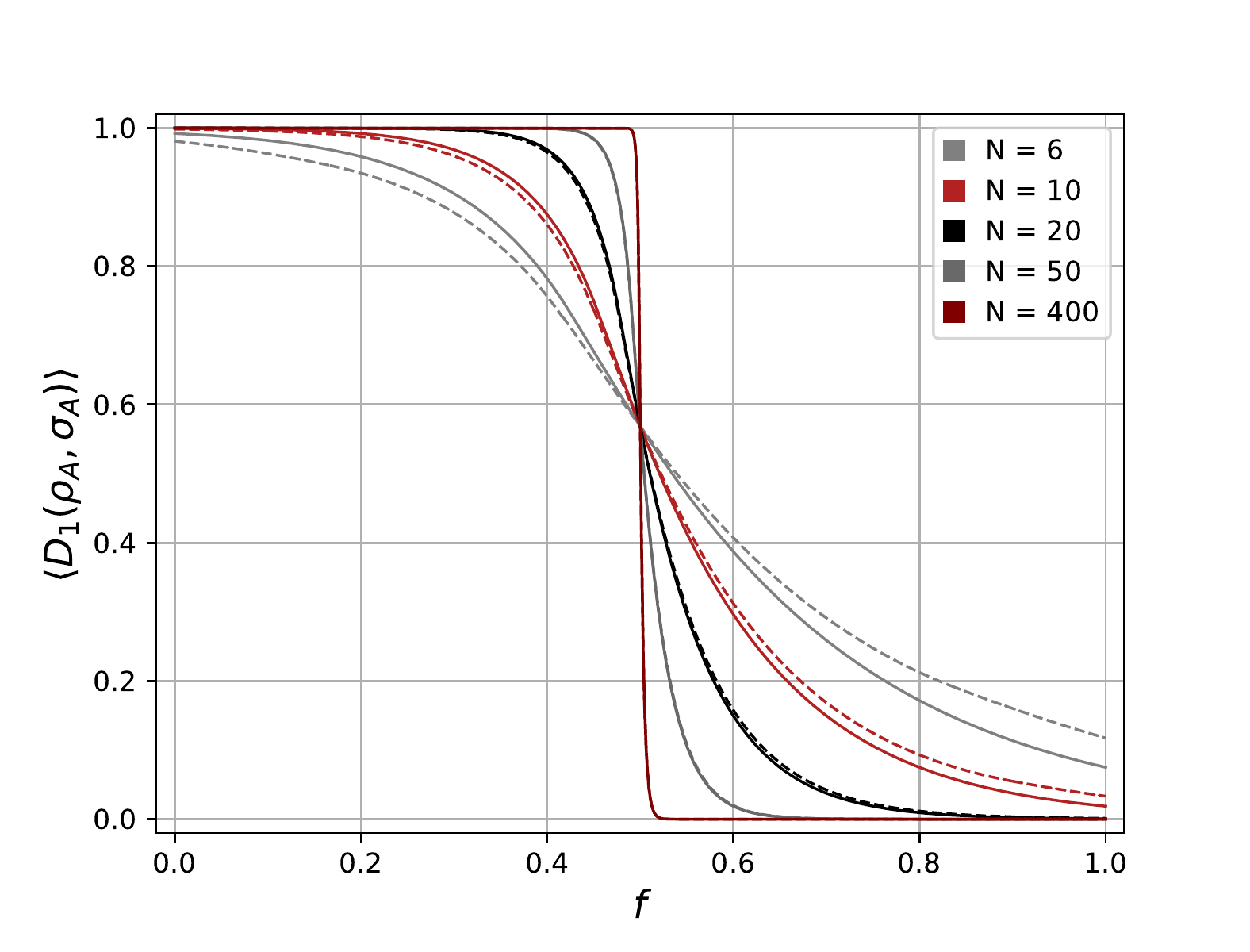}
\vspace{-.5cm}
\caption{
 Average trace-distance of pure random states, Eq.~\eqref{eq:Main_Result},
 (solid lines) and %$Q=0$-
 charge eigenstates for $Q=0$, Eq.~\eqref{eq:Q_eigenstate}, (dashed lines)
 as functions of the relative number of traced qubits $f\equiv N_B/N$.  
System sizes are $N = 6, 10, 20, 50, 400$. 
}
\label{fig2}
\end{figure}
%%%%%%%%%%%%%%%%%%%%%%%%%%%%%%%%%%%%%%%%%%%%%%%%%%%%%%%%

Fig.~\ref{fig2} shows the trace-distance between two random charge eigenstates states, 
Eq.~\eqref{eq:Q_eigenstate}, for different numbers of qubits $N$ (dashed lines)
as a function of the fraction  $f\equiv N_B/N$ of partially traced qubits. 
Differences between trace-distances of charge eigenstates and Page states are notable 
for small systems sizes. They are most pronounced for $f\to 1/N$ and $f\to 1-1/N$, 
with charge eigenstates  
slightly above, respectively, below that of structureless Page states~\cite{footnote6}. 
At half-partition $f=1/2$, $x_f=x$ and ${\cal G}(x,f)={\cal F}(x)$, 
and trace-distances of eigenstates of charges with largest spectral weight 
and Page states are thus identical. This contrasts the entanglement entropy, for which the 
largest difference between Page and  corresponding charge eigenstates 
are at half partition~\cite{Bianchi2019,Lau2020,MonteiroPRL2021}. 

Turning to eigenstates at finite charges $Q$ different 
from the value of largest spectral weight, 
we focus on trace-distances at
half-partition $f=1/2$ and the limits 
$f\to 1/N$, respectively, $f\to 1- 1/N$. For $f=1/2$ 
we find a weak non-monotonous $Q$-dependence of the average trace-distance of charge eigenstates. 
Starting at the value $\langle D_1 \rangle\simeq 0.57$ at $Q=0$, 
it increases to a maximum value $\langle D_1\rangle\simeq 0.58$ 
at $Q=\gamma \sqrt{N}$, before converging to the value $\langle D_1\rangle\simeq 0.50$ as $Q$ is further increased. 
In the limit 
$f\to 1/N$ we find that the average trace-distances decreases with $Q$ as
$\langle D_1 \rangle - 1 \sim -(\sqrt{N}/D)e^{Q^2/2\gamma^2N}$, 
while the leading $Q$-dependence for  
$f\to 1-1/N$ is given by
$\langle D_1\rangle\sim (N^{1/4}/\sqrt{D})e^{Q^2/4\gamma^2N}$. 
In both limits this corresponds to a substitution of $D_S\mapsto F_S(Q)$ 
in the result for Page states,  
accounting for the reduced phase space volume of charge eigenstates 
(see Appendix~\ref{app_charge_eigenstates} for more detailed expression).

\subsection*{Numerical analysis}
\label{sec:exact_diagonalization}

\begin{figure*}[t!]
    \centering
    \includegraphics[width=8.7cm]{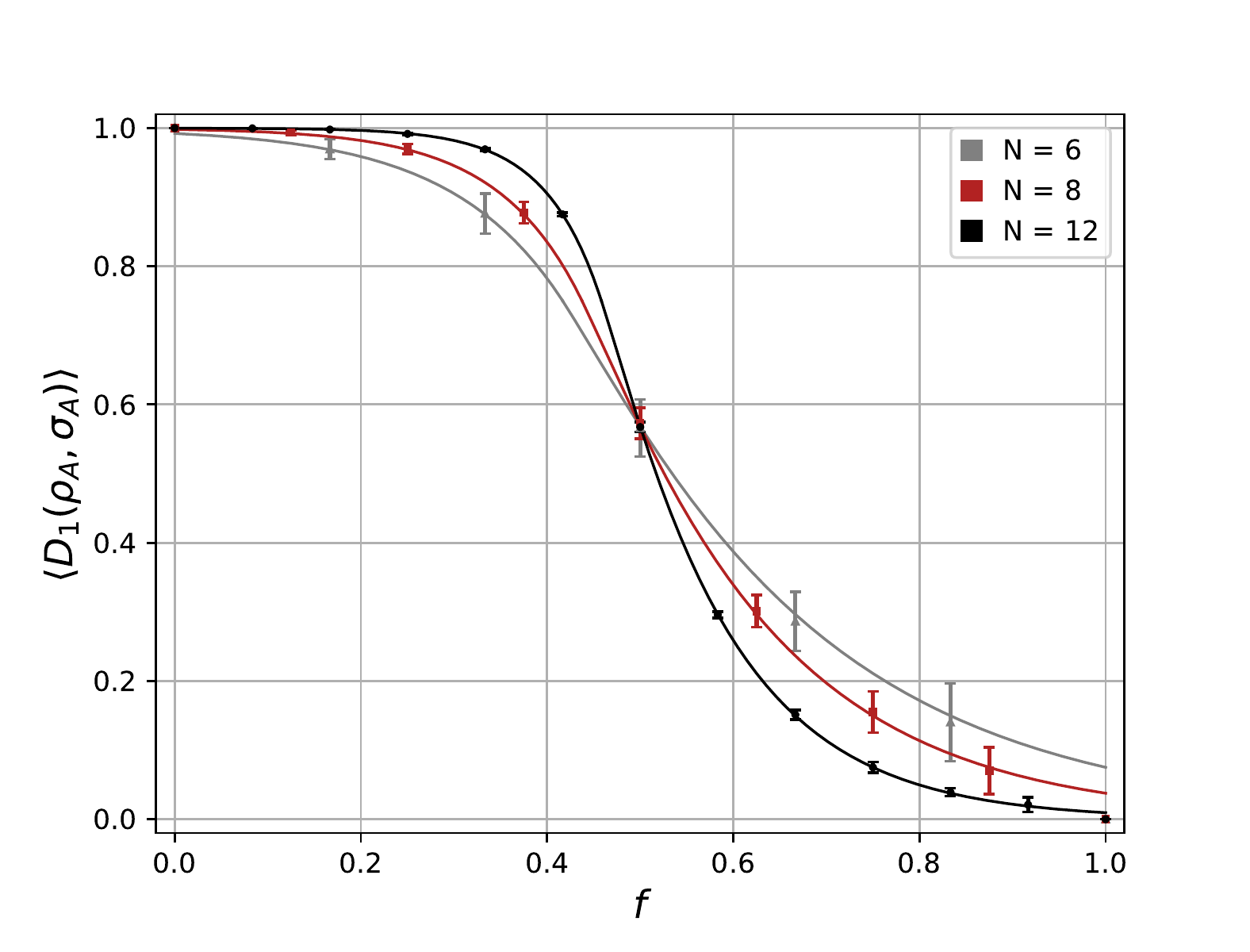}
    \includegraphics[width=8.7cm]{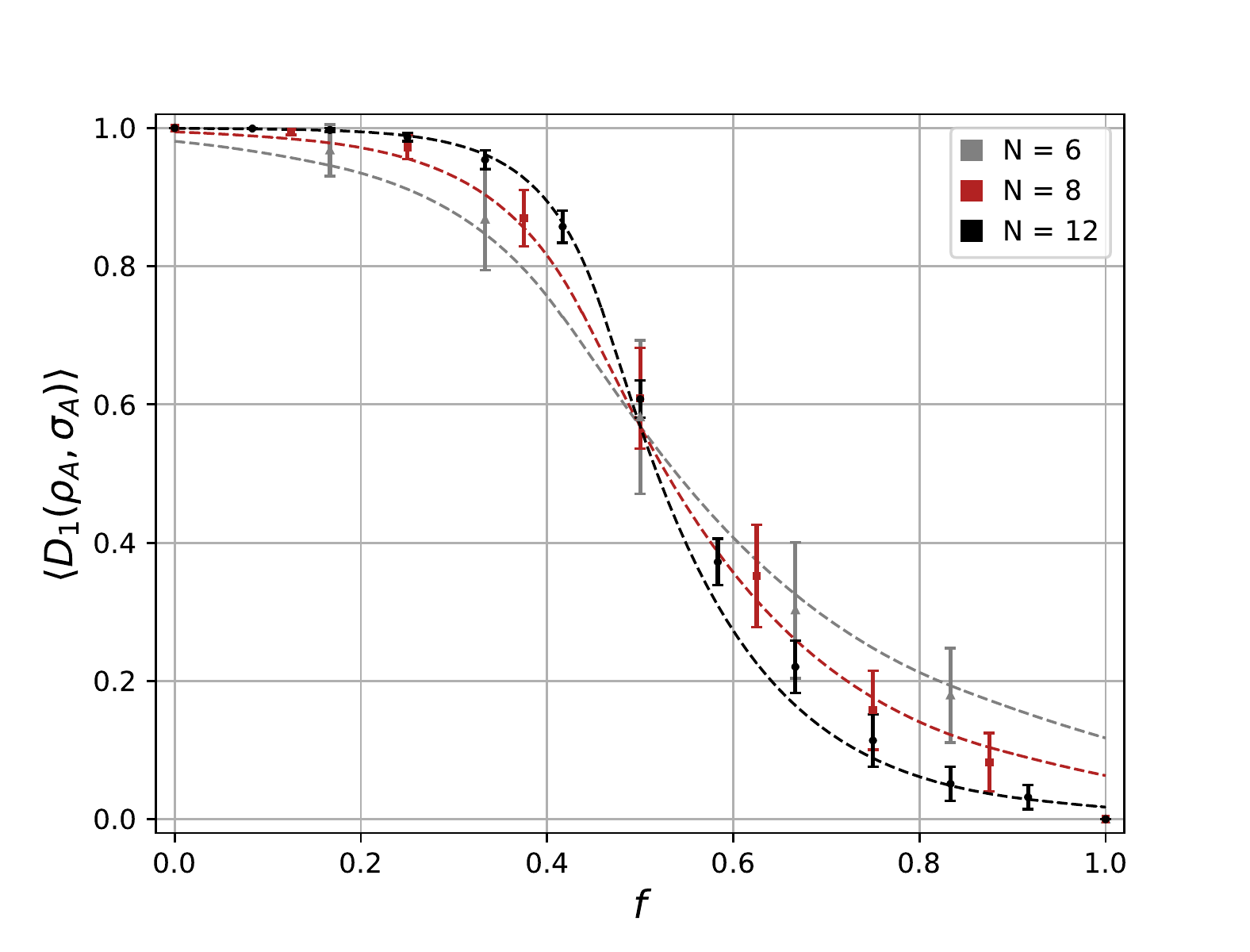}
    \vspace{-4mm}
    \caption{
    Left: Comparison of analytical prediction Eq.~\eqref{eq:Main_Result} (solid lines) and 
 numerical simulations of the SYK model (symbols), here for systems 
 of $N = 14, 18$ and $26$ Majorana fermions, corresponding to $N = 6, 8$ and $12$ qubit systems, 
 viz., the fermion occupations in even parity sector.
   Right: Comparison of analytical prediction Eq.~\eqref{eq:Q_eigenstate} (dashed lines) and 
 numerical simulations of the spin-$1/2$ Ising chain discussed in the main text (symbols), here for  
  $N = 6$, $8$ and $12$ spins. 
 Error bars indicate standard deviation, and for deviations at $f=1$, where $\langle D_1\rangle=0$, 
 see discussion in main text.
    \label{fig3}}
\end{figure*}

Fig.~\ref{fig3} shows a comparison of the analytical predictions Eqs.~\eqref{eq:Main_Result} 
and~\eqref{eq:Q_eigenstate} with numerical results obtained from exact diagonalization of 
Hamiltonians generating many-body chaos. 
The left panel shows eigenstates of an Sachdev-Ye-Kitaev (SYK) model with all-to-all interaction~\cite{SachdevYe1993,KitaevSYK}, 
$\hat{H}_{\rm SYK}
=
\frac{1}{4!}\sum_{i,j,k,l=1}^{N} J_{ijkl} \hat{\chi}_i \hat{\chi}_j \hat{\chi}_k
\hat{\chi}_l $, 
and the right pannel of a spin-$1/2$ Ising chain with nearest neighbor 
interaction and longitudinal and transversal fields~\cite{Huse2013,Huse2014,Huse2015},
$\hat{H}_S=\sum_{i=1}^N \left(g\hat\sigma_i^x + h\hat\sigma_i^z
+ J\hat\sigma_i^z\hat\sigma_{i+1}^z \right)$, and periodic boundary condition $\sigma_{N+1}^z=\sigma_1^z$.
Here $\{\hat \chi_l\}$ are Majorana operators and $\{\hat{\sigma}_i^x, \hat{\sigma}_i^z\}$ Pauli matrices. 
In both cases we have chosen eigenstates from the center of the band, 
to calculate their subsystem trace distances~\cite{footnote8}. 
In the SYK-model we average over $50$ realizations 
of  couplings $J_{ijkl}$ 
(respectively $10$ for the largest system size), 
randomly drawn from a Gaussian distribution 
with vanishing mean and variance $\langle|J_{ijkl}|^2\rangle= 6J^2/N^3$ where we set $J=2/\sqrt{N}$. 
For the spin chain we follow Ref.~\cite{Huse2013,Huse2014,Huse2015}, and use parameters 
$(g,h,J)=(0.9045,0.8090,1.0)$ for which the sytem has been shown to be thermalizing for small 
system sizes. Since  $\hat H_S$ is translational invariant, we first block diagonalize and 
then  average over $7$ eigenstates from a given momentum sector with energies near the band center, 
see also Appendix~\ref{app_zero_spin_chain_simulations} for further details.  

The SYK model lacks local conservation laws, and we find excellent agreement with 
Eq.~\eqref{eq:Main_Result} for Page states. 
For the spin chain with short-range interaction, on the other hand, energy is locally conserved 
and has to be taken into account as a locally conserved charge.
Notice that translational invariance also implies conservation of momentum. This is, however,
not subsystem additive and thus does not count as a locally conserved charge. Rather, 
we restrict 
to a given momentum sector, as described above, and then find good agreement with
 Eq.~\eqref{eq:Q_eigenstate} for systems with a single conserved charge that is subsystem additive.
  Deviations from analytical predictions are larger for the spin chain, 
 which we relate to 
 eigenstates in the average that are not at energies with largest spectral weight  
and deviations of the density of states from Eq.~\eqref{eq:gaussian_spectral_density}. 
Overall the agreement with our analytical predictions for subsystem trace distances 
in absence and presence of locally conserved 
scalar charges, Eqs.~\eqref{eq:Main_Result} and~\eqref{eq:Q_eigenstate}, is very good 
even for the smallest systems with Hilbert-space dimensions $D= 2^6$.

%%%%%%%%%%%%%%%%%%%%%%%%%%%%%%%%%%%%%%%%%%%%%%%%%%%%%%%%
%%%%%%%%%%%%%%%%%%%%%%%%%%%%%%%%%%%%%%%%%%%%%%%%%%%%%%%%

\section{Summary and Discussion}
\label{sec5}

We have studied two-state discrimination in chaotic quantum systems. 
Assuming that one of two generic $N$-qubit random pure states $\rho$, $\sigma$  
has been randomly selected, we investigated the average 
two-state discrimination 
probability $P_{\rho\sigma}=\frac{1}{2}\left(1 +D_1(\rho_A,\sigma_A)\right)$ 
that the selected state is correctly identified from
an optimally chosen experiment on $N_A=N-N_B$ of the qubits. 
Here $D_1(\rho_A,\sigma_A)$ are the subsystem 
trace distances of random pure states with $N_B$ of the qubits partially 
traced out. 
In the thermodynamic limit $N\to\infty$,  
the latter makes a sharp, first order transition from unity to zero 
at $f=1/2$, 
as the fraction $f=N_B/N$ of unmeasured qubits is increased. 
We have given closed analytic expression for the 
corresponding crossover at finite 
system sizes $N$, valid up to corrections small in $1/D$.

We further studied the consequences of local conservation laws on two-state discrimination. 
Specifically, we calculated the average 
subsystem trace-distances for eigenstates of a single conserved scalar 
charge. Focusing on the charge $Q=0$ with largest spectral weight we obtained a closed universal 
expression for the trace distance  that only depends on the numbers of qubits $N$, $N_B$.
We found that state discrimination at the specific point $f=1/2$, i.e. 
involving measurements on half of the qubits, is
 not affected by local conservation laws and the same as for Page states. 
Away from this point, state discrimination in the presence of conservation laws 
becomes less/more likely than in absence of the latter  
for measurements involving more/less than half of the qubits. 
For eigenstates of general charges $Q$, state discrimination at $f=1/2$ shows a weak non-monotonous 
 behavior as $Q$ is varied, while the behavior for different values of $f$  
is similar as for $Q=0$, as we 
 checked in the limits 
$f\to 1/N$ and $1-1/N$, respectively. 
 In all cases, the probability for two-state discrimination 
involving local measurements on non-extensive fractions of the 
$N$ qubits is 1/2 up to corrections $1/D$ for Page states, 
respectively, $1/F(Q)$ for charge eigenstates. 
The exponentially small trace-distance for $f\sim {\cal O}(1)$  
is expected from 
Eigenstate Thermalization Hypothesis (ETH)~\cite{Srednicki1994,Rigol2008,Kaufman2016},
and also in accordance with subsystem ETH, 
its refined formulation  applicable if the entire reduced density matrix appears thermal~\cite{Dymarsky2018}. 
%While the manuscript focuses on the trace distance due to its direct relation with state discrimination, there are other well known measures of distinguishability between quantum states, like the relative entropy (Kullback-Leibler divergence),  the fidelity
%and the associated Bures distance, and the Renyi relative entropies, which set both lower and upper bounds on the trace distance. A comment on these other measures and on the possible implications of these results and methodology  for their behavior and evaluation in the cases considered would be appropriate....
The trace distance bounds other measures for distinguishability of quantum states, such as 
the relative entropy and fidelity via Pinsker's and Fuchs-van de Graaf's inequalities, respectively. 
Similar results e.g. for the consequences of conservation laws on the latter may thus be expected.

We tested our predictions against 
exact diagonalization of Hamiltonians for many-body chaotic systems. 
Specifically, we numerically calculated subsystem trace distances of eigenstates
of the SYK model, lacking any local conservation laws, and  
an Ising spin chain with local energy conservation, respectively, 
and found in both cases very good agreement with our analytial predictions.

We here focused on charge eigenstates and generalizations 
to other pure states in chaotic systems with locally conserved charges~\cite{Sugiura18,Sugiura12,Sugiura13} should 
be interesting.
Based on recent work~\cite{AltlandHuseMicklitz2022},  
we expect that pure states conditioned by broad charge distributions
can be discriminated by local measurements 
(i.e. with probability not exponentially small in $N$)
even in the presence of strong information scrambling. 
Finally, our results may be 
interesting in the context of the black hole information paradox. 
Specifically, one may verify whether the analogy between fixed-area states 
and random tensor networks, encountered for the 
entanglement entropy~\cite{Stanford2019}, continues to hold 
for two-state discrimination of black hole micro-states.

\emph{Acknowledgments:---}We thank Fernando de Mello for discussions. 
T.~M.~acknowledges collaborations with Alex Altland and David Huse on related topics, 
and financial support by Brazilian agencies CNPq and FAPERJ. J.~T.~M.~acknowledges
financial support by Brazilian agency CAPES.

\begin{appendix}

\section{Summing Kreweras numbers}
\label{app_Kreweras}

{\it From Kreweras to Narayanas:---}It is instructive to first 
review how Narayana numbers result from summing Kreweras numbers $\Lambda_n$
with a fixed number of cycles $l(\Lambda_n)=k$.
Substituting the explicit expression discussed in the main text, we can organize this counting as 
\begin{align}
N(n, k) 
&= 
\sum_{\lambda_1=0}^\infty \sum_{\lambda_2=0}^\infty \cdot\cdot\cdot \sum_{\lambda_n=0}^\infty 
\frac{n!\,\delta_{\sum_{i=1}^n \lambda_i, k} 
\,\delta_{\sum_{i=1}^n i\lambda_i, n}}{(n-k+1)!\lambda_1!\lambda_2!...\lambda_n!}, 
\end{align}
where the two Kronecker-deltas fix the number of cycles to $k$ and total  number of elements to $n$, respectively. 
Implementing the latter in terms of integrals $\delta_{x, n} = \frac{1}{2\pi}\int_0^{2\pi} e^{i(x-n)\phi}d\phi$, 
we exchange integration and summation and arrive at,
\begin{align}
N(n, k) 
&=  
\frac{n!}{(n-k+1)!}
\oint \frac{dz}{2\pi i}
\oint\frac{dw}{2\pi i} \frac{e^{z(w+w^2+...+w^n)}}{z^{k+1}w^{n+1}},
\end{align}
which, performing pole integrals gives
\begin{align}
N(n, k) &= 
\frac{\left(\partial_w^n(w+w^2+...+w^n)^k\right)|_{w=0}}{k!(n-k+1)!}.
\end{align}
Summing the finite geometric series one then arrives, 
upon performing the $n$ fold derivative and setting $w=0$, 
at the Narayana numbers
\begin{align}
N(n, k) 
&= 
\frac{1}{n}\binom{n}{k}\binom{n}{k-1}.
\end{align}

{\it From Kreweras to `even-element' Narayanas:---}We can now extend the 
calculation to Kreweras numbers
 $\Lambda_n=(1^0,2^{\lambda_2},3^0...,n^{\lambda_n})$, 
for which  all cycles are composed of even elements, 
\begin{align}
N_e(n, k) 
&= 
\sum_{\lambda_2=0}^\infty \sum_{\lambda_4=0}^\infty \cdot\cdot\cdot \sum_{\lambda_n=0}^\infty 
\frac{n!\,\delta_{\sum_{i=1}^n \lambda_i, k} 
\,\delta_{\sum_{i=1}^n i\lambda_i, n}}{(n-k+1)!\lambda_2!\lambda_4!...\lambda_n!}, 
\end{align}
where $n$ is even. Proceeding then as previously, we find 
\begin{align}
N_e(n, k) 
&=  
\frac{n!}{(n-k+1)!}
\oint \frac{dz}{2\pi i}
\oint\frac{dw}{2\pi i} \frac{e^{z(w^2+w^4+...+w^n)}}{z^{k+1}w^{n+1}},
\end{align}
which performing pole integrals gives,
\begin{align}
N_e(n, k) &= 
\frac{\left(\partial_w^n(w^2+w^4+...+w^n)^k\right)|_{w=0}}{k!(n-k+1)!}.
\end{align}
Summing again the finite geometric series, 
performing the $n$ 
fold derivative and setting $w=0$, 
one then arrives at the `even-element' Narayana numbers, 
\begin{align}
N_e(n, k) 
&= 
\frac{2}{n}\binom{n/2}{k}\binom{n}{k-1},
\end{align}
 stated in the main text.

\section{Subsystem trace-distance}
\label{app_subsystem_trace_distance}

Re-organizing the expression in the main text we find for even $n$-Schatten
distances 
\begin{align}
\label{app_Schantten_n_xleq1}
\langle D_{n}(\rho_A,\sigma_A) \rangle
&= 
\frac{D_A}{nD_B^n}
\sum_{k=1}^\infty 
\binom{n/2}{k}\binom{n}{k-1}
\left(\frac{2D_B}{D_A}\right)^k,
\end{align}  
where we extended the summation to infinity since the binomial restricts $k\leq n/2$.
For $x\equiv 2D_B/D_A\leq 1$ the sum is the hypergeometric function
 ${\cal F}(x)={}_2F_1\left( 1-\frac{n}{2},-n,2,x\right)$ which 
 can be generalized to real $n$. Taking then the replica limit $n\to 1$, 
 \begin{align}
 \langle D_{n}(\rho_A,\sigma_A) \rangle
 &=
 {}_2F_1\left( \frac{1}{2},-1,2,x\right)
=
1-\frac{1}{4x},
 \end{align}
and we recall that $x\leq 1$.
For the complementary case $x> 1$, we 
need to first reorganize the sum \eqref{app_Schantten_n_xleq1},
changing $k\mapsto n/2-k+1$,
\begin{align}
\label{app_Schantten_n_xgeq1}
&\langle D_{n}(\rho_A,\sigma_A) \rangle
= 
\frac{2^{\frac{n}{2}}D_A}{nD^{\frac{n}{2}}}
\sum_{k=0}^\infty 
\binom{n/2}{k}\binom{n}{n/2+1+k}
\left(\frac{D_A}{2D_B}\right)^k,
\end{align}  
where we employed that $\binom{n}{k}=\binom{n}{n-k}$. 
The sum defines 
the hypergeometric function
${\cal F}(x)={}_2F_1\left( -\frac{n}{2},1-\frac{n}{2},2+\frac{n}{2},x\right)$, 
with 
$x=\frac{D_A}{2D_B}$ and can be extended to real $n$. 
Taking the replica limit $n\to 1$ we then arrive 
at Eq.~\eqref{eq:Main_Result} in the main text.

\section{Charge eigenstates}
\label{app_charge_eigenstates}

Straightforward generalization to charge eigenstates defined in the main text,
we arrive at
\begin{align}
\label{app_trace_distance_charge_eigenstates}
\langle D_1&(\rho_A,\sigma_A) 
\rangle
= 
\frac{1}{F(Q)}
\sum_{Q_AQ_B}\delta_{Q,Q_A+Q_B}
\Bigg\{
\nonumber\\
&\quad
  \left(
  F_A(Q_A)F_B(Q_B)
  -\frac{1}{2}F_B^2(Q_B)
  \right)\Theta_<
  \nonumber\\
  &+
  \frac{4\sqrt{2}}{3\pi}
F_A^{3/2}(Q_A)F^{1/2}_B(Q_B)
{}_2F_1\left( -\frac{1}{2},\frac{1}{2},\frac{5}{2},x\right)
\Theta_>
  \Bigg\},
\end{align}
where $\Theta_{<}\equiv
\theta\left(x-1\right)
$,
$\Theta_{>}\equiv
\theta\left(1-x\right)
$, with
$x\equiv F_A(Q_A)/(2F_B(Q_B))$, 
$F_S(Q)=D_S\Omega_S(Q)$, and 
$\Omega_S(Q)=\frac{1}{\sqrt{2\pi\gamma^2N_S}}\exp\left(-Q_S^2/(2\gamma^2 N_S)\right)$.

{\it $Q=0$:---}Concentrating first on 
the charge with largest spectral weight 
$Q=0$ where the density of states is peaked, 
we can 
substitute ($N_A$, $N_B$ are integers), 
$\Theta_<=\theta(N_A-N_B)$
and $\Theta_>=\theta(N_B-N_A)$.
Using further that
\begin{align}
&\frac{1}{F(0)}
\sum_{Q_A}
F_A^k(Q_A)F_B^m(-Q_A)
\nonumber\\
&=
\frac{D^k_AD^m_B}{(2\pi\gamma^2)^{\frac{k+m}{2}-1}D}
\sqrt{\frac{N_AN_B}{N_A^kN^m_B}}
\sqrt{\frac{N}{mN_A+kN_B}},
\end{align}
we arrive at,
\begin{align}
\label{app:Q_eigenstate}
&\langle 
D_1(\rho_A, \sigma_A) 
\rangle
\nonumber\\
&= 
\begin{cases}
1-\frac{1}{4x\sqrt{2f}}
&x\geq 1,
\\
\sum_{k=0}^\infty
\frac{2\sqrt{x_f} x_f^k}{\sqrt{(1+2k)f+(1/2-k)}}
\binom{1/2}{k}\binom{1}{k+3/2}, 
&x \leq 1,
\end{cases}
\end{align}
where 
$x_f \equiv 
x\sqrt{f/(1 - f)}$, $x=D_A/(2D_B)$, and $f = N_B/N$, as stated in 
Eq.~\eqref{eq:Q_eigenstate} in the main text.

{\it Finite charges:---}Expressions for finite charges $Q>0$ can be derived in 
a similar way. We here concentrate on half partitions $N_A=N_B=N/2$, and the limits 
$f\to 1/N$, 
respectively, $f\to 1-1/N$. 
For half partitions 
$N_A=N_B=N/2$, we can use that
\begin{align}
\frac{1}{F(Q)}
&\sum_{Q_A}
F_A^k(Q_A)F_B^m(Q-Q_A)\theta_S(Q_A)
\nonumber\\
=&
\frac{D_A^kD_B^m}{D}
\frac{\sqrt{2\pi\gamma^2N}}{(\pi\gamma^2N)^{\frac{k+m}{2}}}
e^{-\frac{2km-k-m}{k+m}\frac{Q^2}{2\gamma^2 N}}
\times
\nonumber\\
&\times\sum_{Q_A}
e^{-\frac{(k+m)Q_A^2}{\gamma^2 N}}\theta_S(Q_A+\frac{m}{k+m}Q),
\end{align}
where 
$\theta_<(Q_A+\frac{m}{k+m}Q)\equiv\theta(Q/2-\frac{NC}{2Q}-\frac{m}{k+m}Q-Q_A)$, 
$\theta_>(Q_A+\frac{m}{k+m}Q)\equiv\theta(Q_A-Q/2+\frac{NC}{2Q}+\frac{m}{k+m}Q)$, 
and $C\equiv\gamma^2\ln2$.
With this we then arrive at the following expression for the trace-distance at half partition,
\begin{align}
\langle D_1(\rho_A,& \sigma_A)\rangle = \frac{1}{2}\erfc\Bigg(\sqrt{\frac{N}{2\gamma^2}}\frac{C}{Q}\Bigg)
\nonumber\\
&-\frac{1}{4}e^{\frac{Q^2}{2\gamma^2N}}\erfc\Bigg(\frac{NC+Q^2}{Q\sqrt{2\gamma^2N}}\Bigg)
\nonumber\\
&+\sum_{k=0}^\infty\binom{1/2}{k}\binom{1}{k+3/2}\Bigg[\frac{e^{
\frac{(k+1/2)Q^2}{2\gamma^2N}}}{2}\Bigg]^{k+1/2}\times
\nonumber\\
&\times\erfc\Bigg(\frac{(1+2k)Q^2-2NC}{2Q\sqrt{2\gamma^2N}}\Bigg),
\end{align}
where $\erfc(x)=(2/\sqrt{\pi})\int_x^\infty e^{-t^2}dt$. 
This can be evaluated numerically and shows the $Q$-dependence discussed in the main text.

For $f\to 1/N$  
we can neglect the contribution involving $\Theta_>$, 
and find 
$\langle D_1 \rangle -1  \sim -\sqrt{N}e^{Q^2/2\gamma^2N}/D$. 
Proceeding similarly in the opposite limit $f\to 1-1/N$,  
we arrive at
$\langle D_1\rangle\sim N^{1/4}e^{Q^2/4\gamma^2N}/\sqrt{D}$, 
as also stated in the main text. 

%%%%%%%%%%%%%%%%%%%%%%%%%%%%%%%%%%%%%%%%%%%%%%%%%%%%%%%%
\begin{figure}[h!]
    \centering
    \includegraphics[width=8.7cm]{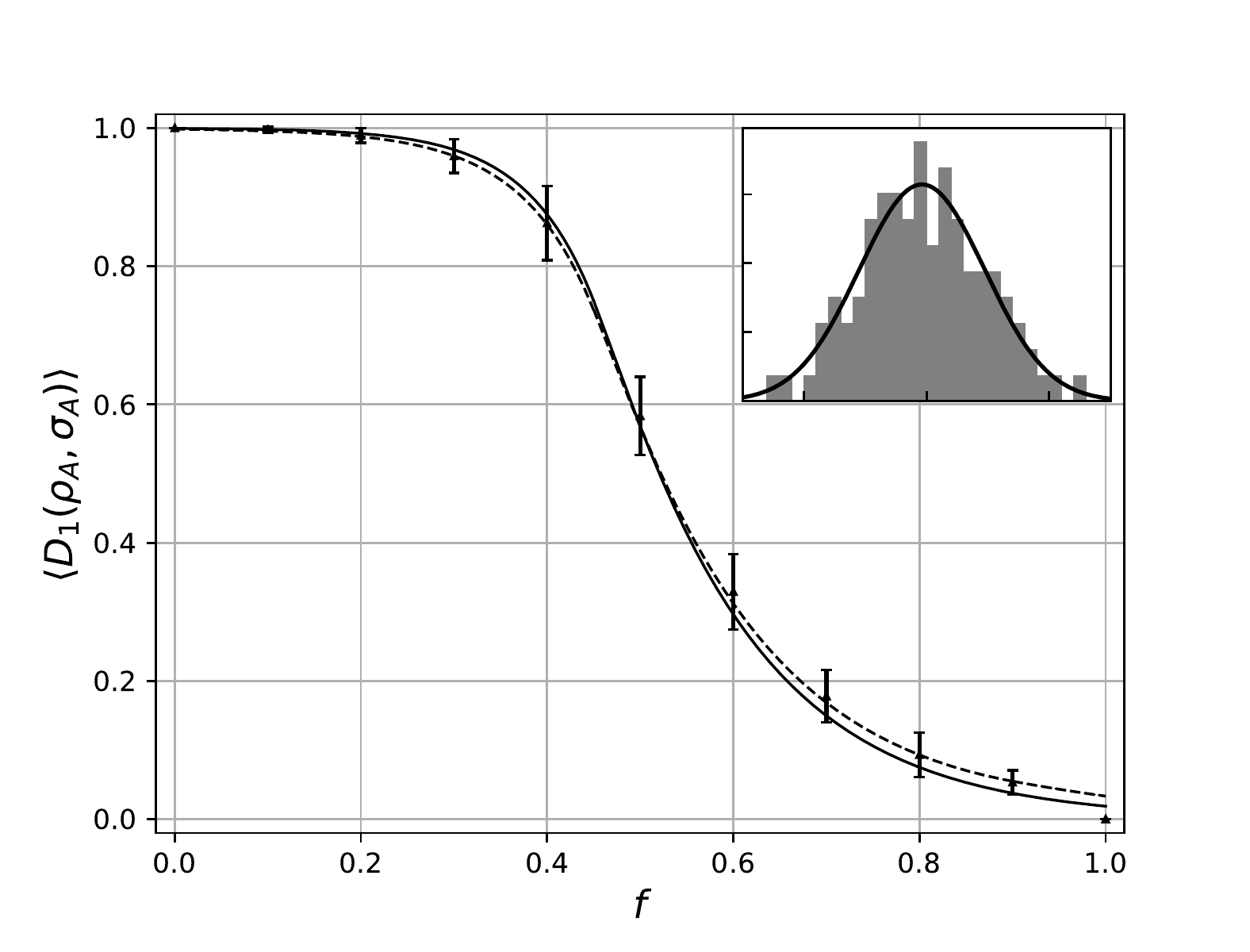}
    \vspace{-4mm}
    \caption{
Subsystem trace-distances from exact diagoanlization for a chain of $10$ spins. Solid and dashed lines are the analytical predictions in 
 absence and presence of conservation laws, Eqs.~\eqref{eq:Main_Result} and Eq.~\eqref{eq:Q_eigenstate}, respectively. 
    Inset: Density of states for zero momentum eigenstates with  Gaussian fit (solid line).}
    \label{fig4}
\end{figure}
%%%%%

\section{Exact diagonaliztion}
\label{app_zero_spin_chain_simulations}

We numerically calculate eigenstates of the spin chain Hamiltonian first block diagonalizing $\hat H_S$ into momentum sectors, and then concentrating on eigenfunctions within 
the zero momentum eigenspace. To determine the energy window from which to choose eigenstates we calculate the density of states, 
see inset of Fig.~\ref{fig4} for the example of the chain with 10 spins. 
The latter is e.g. peaked at $-0.4 \pm 0.3$ and reasonably well described by a Gaussian profile (solid line). 
Taking in this case $5$ eigenstates from the window $\sim (-0.8, 0.0)$ centered around $\sim -0.4$
we arrive at the subsystem trace distance shown in Fig.~\ref{fig4}, and 
 compared to the analytical prediction in presence (dashed line) and absence (solid line) of a
local conservation law.

\end{appendix}


\begin{thebibliography}{36}
\expandafter\ifx\csname natexlab\endcsname\relax\def\natexlab#1{#1}\fi
\expandafter\ifx\csname bibnamefont\endcsname\relax
  \def\bibnamefont#1{#1}\fi
\expandafter\ifx\csname bibfnamefont\endcsname\relax
  \def\bibfnamefont#1{#1}\fi
\expandafter\ifx\csname citenamefont\endcsname\relax
  \def\citenamefont#1{#1}\fi
\expandafter\ifx\csname url\endcsname\relax
  \def\url#1{\texttt{#1}}\fi
\expandafter\ifx\csname urlprefix\endcsname\relax\def\urlprefix{URL }\fi
\providecommand{\bibinfo}[2]{#2}
\providecommand{\eprint}[2][]{\url{#2}}

\bibitem[{\citenamefont{Page}(1993)}]{Page1993}
\bibinfo{author}{\bibfnamefont{D.~N.} \bibnamefont{Page}},
  \bibinfo{journal}{Phys. Rev. Lett.} \textbf{\bibinfo{volume}{71}},
  \bibinfo{pages}{1291} (\bibinfo{year}{1993}),
  \urlprefix\url{https://link.aps.org/doi/10.1103/PhysRevLett.71.1291}.

\bibitem[{\citenamefont{Bianchi and Don\`a}(2019)}]{Bianchi2019}
\bibinfo{author}{\bibfnamefont{E.}~\bibnamefont{Bianchi}} \bibnamefont{and}
  \bibinfo{author}{\bibfnamefont{P.}~\bibnamefont{Don\`a}},
  \bibinfo{journal}{Phys. Rev. D} \textbf{\bibinfo{volume}{100}},
  \bibinfo{pages}{105010} (\bibinfo{year}{2019}),
  \urlprefix\url{https://link.aps.org/doi/10.1103/PhysRevD.100.105010}.

\bibitem[{\citenamefont{Bianchi et~al.}(2021)\citenamefont{Bianchi, Hackl,
  Kieburg, Rigol, and Vidmar}}]{RigolKieburg2021}
\bibinfo{author}{\bibfnamefont{E.}~\bibnamefont{Bianchi}},
  \bibinfo{author}{\bibfnamefont{L.}~\bibnamefont{Hackl}},
  \bibinfo{author}{\bibfnamefont{M.}~\bibnamefont{Kieburg}},
  \bibinfo{author}{\bibfnamefont{M.}~\bibnamefont{Rigol}}, \bibnamefont{and}
  \bibinfo{author}{\bibfnamefont{L.}~\bibnamefont{Vidmar}},
  \emph{\bibinfo{title}{Volume-law entanglement entropy of typical pure quantum
  states}} (\bibinfo{year}{2021}),
  \urlprefix\url{https://arxiv.org/abs/2112.06959}.

\bibitem[{\citenamefont{Rigol et~al.}(2008)\citenamefont{Rigol, Dunjko, and
  Olshanii}}]{Rigol2008}
\bibinfo{author}{\bibfnamefont{M.}~\bibnamefont{Rigol}},
  \bibinfo{author}{\bibfnamefont{V.}~\bibnamefont{Dunjko}}, \bibnamefont{and}
  \bibinfo{author}{\bibfnamefont{M.}~\bibnamefont{Olshanii}},
  \bibinfo{journal}{Nature} \textbf{\bibinfo{volume}{452}},
  \bibinfo{pages}{854} (\bibinfo{year}{2008}),
  \urlprefix\url{https://doi.org/10.1038/nature06838}.

\bibitem[{\citenamefont{Lau et~al.}(2022)\citenamefont{Lau, Noumi, Takii, and
  Tamaoka}}]{Lau2020}
\bibinfo{author}{\bibfnamefont{P.~H.~C.} \bibnamefont{Lau}},
  \bibinfo{author}{\bibfnamefont{T.}~\bibnamefont{Noumi}},
  \bibinfo{author}{\bibfnamefont{Y.}~\bibnamefont{Takii}}, \bibnamefont{and}
  \bibinfo{author}{\bibfnamefont{K.}~\bibnamefont{Tamaoka}},
  \emph{\bibinfo{title}{Page curve and symmetries}} (\bibinfo{year}{2022}),
  \urlprefix\url{https://arxiv.org/abs/2206.09633}.

\bibitem[{\citenamefont{Ares et~al.}(2022)\citenamefont{Ares, Murciano, and
  Calabrese}}]{Ares2022}
\bibinfo{author}{\bibfnamefont{F.}~\bibnamefont{Ares}},
  \bibinfo{author}{\bibfnamefont{S.}~\bibnamefont{Murciano}}, \bibnamefont{and}
  \bibinfo{author}{\bibfnamefont{P.}~\bibnamefont{Calabrese}},
  \bibinfo{journal}{Journal of Statistical Mechanics: Theory and Experiment}
  \textbf{\bibinfo{volume}{2022}}, \bibinfo{pages}{063104}
  (\bibinfo{year}{2022}), \urlprefix\url{https://doi.org/10.1088}.

\bibitem[{\citenamefont{Vidmar and Rigol}(2017)}]{VidmarRigol17}
\bibinfo{author}{\bibfnamefont{L.}~\bibnamefont{Vidmar}} \bibnamefont{and}
  \bibinfo{author}{\bibfnamefont{M.}~\bibnamefont{Rigol}},
  \bibinfo{journal}{Phys. Rev. Lett.} \textbf{\bibinfo{volume}{119}},
  \bibinfo{pages}{220603} (\bibinfo{year}{2017}),
  \urlprefix\url{https://link.aps.org/doi/10.1103/PhysRevLett.119.220603}.

\bibitem[{\citenamefont{Nakagawa et~al.}(2018)\citenamefont{Nakagawa, Watanabe,
  Fujita, and Sugiura}}]{Sugiura18}
\bibinfo{author}{\bibfnamefont{Y.~O.} \bibnamefont{Nakagawa}},
  \bibinfo{author}{\bibfnamefont{M.}~\bibnamefont{Watanabe}},
  \bibinfo{author}{\bibfnamefont{H.}~\bibnamefont{Fujita}}, \bibnamefont{and}
  \bibinfo{author}{\bibfnamefont{S.}~\bibnamefont{Sugiura}},
  \bibinfo{journal}{Nature Communications} \textbf{\bibinfo{volume}{9}},
  \bibinfo{pages}{1635} (\bibinfo{year}{2018}).

\bibitem[{\citenamefont{Sugiura and Shimizu}(2012)}]{Sugiura12}
\bibinfo{author}{\bibfnamefont{S.}~\bibnamefont{Sugiura}} \bibnamefont{and}
  \bibinfo{author}{\bibfnamefont{A.}~\bibnamefont{Shimizu}},
  \bibinfo{journal}{Phys. Rev. Lett.} \textbf{\bibinfo{volume}{108}},
  \bibinfo{pages}{240401} (\bibinfo{year}{2012}),
  \urlprefix\url{https://link.aps.org/doi/10.1103/PhysRevLett.108.240401}.

\bibitem[{\citenamefont{Sugiura and Shimizu}(2013)}]{Sugiura13}
\bibinfo{author}{\bibfnamefont{S.}~\bibnamefont{Sugiura}} \bibnamefont{and}
  \bibinfo{author}{\bibfnamefont{A.}~\bibnamefont{Shimizu}},
  \bibinfo{journal}{Phys. Rev. Lett.} \textbf{\bibinfo{volume}{111}},
  \bibinfo{pages}{010401} (\bibinfo{year}{2013}),
  \urlprefix\url{https://link.aps.org/doi/10.1103/PhysRevLett.111.010401}.

\bibitem[{\citenamefont{Murciano et~al.}(2022)\citenamefont{Murciano,
  Calabrese, and Piroli}}]{Calabrese2022}
\bibinfo{author}{\bibfnamefont{S.}~\bibnamefont{Murciano}},
  \bibinfo{author}{\bibfnamefont{P.}~\bibnamefont{Calabrese}},
  \bibnamefont{and} \bibinfo{author}{\bibfnamefont{L.}~\bibnamefont{Piroli}},
  \bibinfo{journal}{Phys. Rev. D} \textbf{\bibinfo{volume}{106}},
  \bibinfo{pages}{046015} (\bibinfo{year}{2022}),
  \urlprefix\url{https://link.aps.org/doi/10.1103/PhysRevD.106.046015}.

\bibitem[{\citenamefont{Majidy et~al.}(2022)\citenamefont{Majidy, Lasek, Huse,
  and Halpern}}]{Halpern2022}
\bibinfo{author}{\bibfnamefont{S.}~\bibnamefont{Majidy}},
  \bibinfo{author}{\bibfnamefont{A.}~\bibnamefont{Lasek}},
  \bibinfo{author}{\bibfnamefont{D.~A.} \bibnamefont{Huse}}, \bibnamefont{and}
  \bibinfo{author}{\bibfnamefont{N.~Y.} \bibnamefont{Halpern}},
  \emph{\bibinfo{title}{Non-abelian symmetry can increase entanglement
  entropy}} (\bibinfo{year}{2022}),
  \urlprefix\url{https://arxiv.org/abs/2209.14303}.

\bibitem[{\citenamefont{Altland et~al.}(2022)\citenamefont{Altland, Huse, and
  Micklitz}}]{AltlandHuseMicklitz2022}
\bibinfo{author}{\bibfnamefont{A.}~\bibnamefont{Altland}},
  \bibinfo{author}{\bibfnamefont{D.~A.} \bibnamefont{Huse}}, \bibnamefont{and}
  \bibinfo{author}{\bibfnamefont{T.}~\bibnamefont{Micklitz}},
  \emph{\bibinfo{title}{Maximum entropy quantum state distributions}}
  (\bibinfo{year}{2022}), \urlprefix\url{https://arxiv.org/abs/2203.12580}.

\bibitem[{\citenamefont{Bae and Kwek}(2015)}]{Bae2015}
\bibinfo{author}{\bibfnamefont{J.}~\bibnamefont{Bae}} \bibnamefont{and}
  \bibinfo{author}{\bibfnamefont{L.-C.} \bibnamefont{Kwek}},
  \bibinfo{journal}{Journal of Physics A: Mathematical and Theoretical}
  \textbf{\bibinfo{volume}{48}}, \bibinfo{pages}{083001}
  (\bibinfo{year}{2015}), \urlprefix\url{https://doi.org/10.1088}.

\bibitem[{foo({\natexlab{a}})}]{footnote1}
\bibinfo{note}{There are two kinds of errors i.e. the probability $p_\rho$ one
  guesses wrong when it is $\rho$, and the probability $p_\sigma$ one guess
  wrong when it is $\sigma$, respectively. The best success probability
  minimizes the maximum $\max(p_\rho, p_\sigma)$.}

\bibitem[{foo({\natexlab{b}})}]{footnote2}
\bibinfo{note}{That is, the eigenvalues of $\sqrt{\Lambda^\dagger\Lambda}$, and
  when $\Lambda$ is hermitean, singular values $\lambda_i$ are just the
  absolute values of the eigenvalues.}

\bibitem[{\citenamefont{Zhang et~al.}(2019)\citenamefont{Zhang, Ruggiero, and
  Calabrese}}]{Calabrese2019}
\bibinfo{author}{\bibfnamefont{J.}~\bibnamefont{Zhang}},
  \bibinfo{author}{\bibfnamefont{P.}~\bibnamefont{Ruggiero}}, \bibnamefont{and}
  \bibinfo{author}{\bibfnamefont{P.}~\bibnamefont{Calabrese}},
  \bibinfo{journal}{Phys. Rev. Lett.} \textbf{\bibinfo{volume}{122}},
  \bibinfo{pages}{141602} (\bibinfo{year}{2019}),
  \urlprefix\url{https://link.aps.org/doi/10.1103/PhysRevLett.122.141602}.

\bibitem[{\citenamefont{Monteiro et~al.}(2021)\citenamefont{Monteiro, Tezuka,
  Altland, Huse, and Micklitz}}]{MonteiroPRL2021}
\bibinfo{author}{\bibfnamefont{F.}~\bibnamefont{Monteiro}},
  \bibinfo{author}{\bibfnamefont{M.}~\bibnamefont{Tezuka}},
  \bibinfo{author}{\bibfnamefont{A.}~\bibnamefont{Altland}},
  \bibinfo{author}{\bibfnamefont{D.~A.} \bibnamefont{Huse}}, \bibnamefont{and}
  \bibinfo{author}{\bibfnamefont{T.}~\bibnamefont{Micklitz}},
  \bibinfo{journal}{Phys. Rev. Lett.} \textbf{\bibinfo{volume}{127}},
  \bibinfo{pages}{030601} (\bibinfo{year}{2021}),
  \urlprefix\url{https://link.aps.org/doi/10.1103/PhysRevLett.127.030601}.

\bibitem[{foo({\natexlab{c}})}]{footnote3}
\bibinfo{note}{Application of the replica trick allows for a calculation of the
  entanglement entropy from average moments of the reduced density matrix,
  $M_r=\langle\mathrm{tr}_A(\rho_A^r)\rangle$, as $S_A=-\partial_{r}
  M_r|_{r=1}$.}

\bibitem[{\citenamefont{Penington et~al.}(2019)\citenamefont{Penington,
  Shenker, Stanford, and Yang}}]{Stanford2019}
\bibinfo{author}{\bibfnamefont{G.}~\bibnamefont{Penington}},
  \bibinfo{author}{\bibfnamefont{S.~H.} \bibnamefont{Shenker}},
  \bibinfo{author}{\bibfnamefont{D.}~\bibnamefont{Stanford}}, \bibnamefont{and}
  \bibinfo{author}{\bibfnamefont{Z.}~\bibnamefont{Yang}},
  \emph{\bibinfo{title}{Replica wormholes and the black hole interior}}
  (\bibinfo{year}{2019}), \eprint{arXiv:1911.11977}.

\bibitem[{\citenamefont{Liu and Vardhan}(2020)}]{Liu20}
\bibinfo{author}{\bibfnamefont{H.}~\bibnamefont{Liu}} \bibnamefont{and}
  \bibinfo{author}{\bibfnamefont{S.}~\bibnamefont{Vardhan}},
  \emph{\bibinfo{title}{Entanglement entropies of equilibrated pure states in
  quantum many-body systems and gravity}} (\bibinfo{year}{2020}),
  \eprint{arXiv:2008.01089}.

\bibitem[{\citenamefont{Kreweras}(1972)}]{KREWERAS1972333}
\bibinfo{author}{\bibfnamefont{G.}~\bibnamefont{Kreweras}},
  \bibinfo{journal}{Discrete Mathematics} \textbf{\bibinfo{volume}{1}},
  \bibinfo{pages}{333} (\bibinfo{year}{1972}), ISSN \bibinfo{issn}{0012-365X},
  \urlprefix\url{https://www.sciencedirect.com/science/article/pii/0012365X72900416}.

\bibitem[{\citenamefont{Simion}(2000)}]{SIMION2000367}
\bibinfo{author}{\bibfnamefont{R.}~\bibnamefont{Simion}},
  \bibinfo{journal}{Discrete Mathematics} \textbf{\bibinfo{volume}{217}},
  \bibinfo{pages}{367} (\bibinfo{year}{2000}), ISSN \bibinfo{issn}{0012-365X},
  \urlprefix\url{https://www.sciencedirect.com/science/article/pii/S0012365X99002733}.

\bibitem[{foo({\natexlab{d}})}]{footnote5}
\bibinfo{note}{The simplest way to see this, is to account for the sign factor
  ${\rm sgn}(\sigma)$ in Eq.~(4) by defining averages for states $\sigma$ with
  a minus sign, i.e. $\langle
  \psi^\sigma_{ab}\bar{\psi}^\sigma_{cd}\rangle=-\frac{1}{D}\delta_{ac}\delta_{bd}$.
  Then every $\sigma$ cycle containing odd/even elements contributes with a
  negative/positive sign.}

\bibitem[{\citenamefont{Kudler-Flam et~al.}(2021)\citenamefont{Kudler-Flam,
  Narovlansky, and Ryu}}]{Ryu2021}
\bibinfo{author}{\bibfnamefont{J.}~\bibnamefont{Kudler-Flam}},
  \bibinfo{author}{\bibfnamefont{V.}~\bibnamefont{Narovlansky}},
  \bibnamefont{and} \bibinfo{author}{\bibfnamefont{S.}~\bibnamefont{Ryu}},
  \bibinfo{journal}{PRX Quantum} \textbf{\bibinfo{volume}{2}},
  \bibinfo{pages}{040340} (\bibinfo{year}{2021}),
  \urlprefix\url{https://link.aps.org/doi/10.1103/PRXQuantum.2.040340}.

\bibitem[{\citenamefont{Pucha{\l}a et~al.}(2016)\citenamefont{Pucha{\l}a,
  Pawela, and {\.Z}yczkowski}}]{Zyczkowski2016}
\bibinfo{author}{\bibfnamefont{Z.}~\bibnamefont{Pucha{\l}a}},
  \bibinfo{author}{\bibfnamefont{{\L}.}~\bibnamefont{Pawela}},
  \bibnamefont{and}
  \bibinfo{author}{\bibfnamefont{K.}~\bibnamefont{{\.Z}yczkowski}},
  \bibinfo{journal}{Phys. Rev. A} \textbf{\bibinfo{volume}{93}},
  \bibinfo{pages}{062112} (\bibinfo{year}{2016}),
  \urlprefix\url{https://link.aps.org/doi/10.1103/PhysRevA.93.062112}.

\bibitem[{foo({\natexlab{e}})}]{footnote6}
\bibinfo{note}{Notice that Eq.~(11) diverges in the limits $f=0$ and $f=1$,
  where $F_B$ and $F_A$ become $\delta$-functions. In Fig.~2 we have
  interpolated with a quadratic polynomial to values from Eq.~(C1) (Appendix
  C), with $F_B=1$, $F_A(Q_A)=F(Q)$ and $F_A=1$, $F_ B(Q_B)=F(Q)$,
  respectively, with an error of order ${\cal O}(1/F(0))$.}

\bibitem[{\citenamefont{Sachdev and Ye}(1993)}]{SachdevYe1993}
\bibinfo{author}{\bibfnamefont{S.}~\bibnamefont{Sachdev}} \bibnamefont{and}
  \bibinfo{author}{\bibfnamefont{J.}~\bibnamefont{Ye}}, \bibinfo{journal}{Phys.
  Rev. Lett.} \textbf{\bibinfo{volume}{70}}, \bibinfo{pages}{3339}
  (\bibinfo{year}{1993}),
  \urlprefix\url{https://link.aps.org/doi/10.1103/PhysRevLett.70.3339}.

\bibitem[{Kit()}]{KitaevSYK}
\bibinfo{note}{A.~Kitaev, http://online.kitp.ucsb.edu/online/
  entangled15/kitaev/ .... /kitaev2/ (Talks at KITP on April 7th and May 27th
  2015).}

\bibitem[{\citenamefont{Kim and Huse}(2013)}]{Huse2013}
\bibinfo{author}{\bibfnamefont{H.}~\bibnamefont{Kim}} \bibnamefont{and}
  \bibinfo{author}{\bibfnamefont{D.~A.} \bibnamefont{Huse}},
  \bibinfo{journal}{Phys. Rev. Lett.} \textbf{\bibinfo{volume}{111}},
  \bibinfo{pages}{127205} (\bibinfo{year}{2013}),
  \urlprefix\url{https://link.aps.org/doi/10.1103/PhysRevLett.111.127205}.

\bibitem[{\citenamefont{Kim et~al.}(2014)\citenamefont{Kim, Ikeda, and
  Huse}}]{Huse2014}
\bibinfo{author}{\bibfnamefont{H.}~\bibnamefont{Kim}},
  \bibinfo{author}{\bibfnamefont{T.~N.} \bibnamefont{Ikeda}}, \bibnamefont{and}
  \bibinfo{author}{\bibfnamefont{D.~A.} \bibnamefont{Huse}},
  \bibinfo{journal}{Phys. Rev. E} \textbf{\bibinfo{volume}{90}},
  \bibinfo{pages}{052105} (\bibinfo{year}{2014}),
  \urlprefix\url{https://link.aps.org/doi/10.1103/PhysRevE.90.052105}.

\bibitem[{\citenamefont{Zhang et~al.}(2015)\citenamefont{Zhang, Kim, and
  Huse}}]{Huse2015}
\bibinfo{author}{\bibfnamefont{L.}~\bibnamefont{Zhang}},
  \bibinfo{author}{\bibfnamefont{H.}~\bibnamefont{Kim}}, \bibnamefont{and}
  \bibinfo{author}{\bibfnamefont{D.~A.} \bibnamefont{Huse}},
  \bibinfo{journal}{Phys. Rev. E} \textbf{\bibinfo{volume}{91}},
  \bibinfo{pages}{062128} (\bibinfo{year}{2015}),
  \urlprefix\url{https://link.aps.org/doi/10.1103/PhysRevE.91.062128}.

\bibitem[{foo({\natexlab{f}})}]{footnote8}
\bibinfo{note}{The SYK model preserves fermion parity and (without loss of
  generality) we have chosen eigenstates from the even parity sector.}

\bibitem[{\citenamefont{Srednicki}(1994)}]{Srednicki1994}
\bibinfo{author}{\bibfnamefont{M.}~\bibnamefont{Srednicki}},
  \bibinfo{journal}{Phys. Rev. E} \textbf{\bibinfo{volume}{50}},
  \bibinfo{pages}{888} (\bibinfo{year}{1994}),
  \urlprefix\url{https://link.aps.org/doi/10.1103/PhysRevE.50.888}.

\bibitem[{\citenamefont{Kaufman et~al.}(2016)\citenamefont{Kaufman, Tai, Lukin,
  Rispoli, Schittko, Preiss, and Greiner}}]{Kaufman2016}
\bibinfo{author}{\bibfnamefont{A.~M.} \bibnamefont{Kaufman}},
  \bibinfo{author}{\bibfnamefont{M.~E.} \bibnamefont{Tai}},
  \bibinfo{author}{\bibfnamefont{A.}~\bibnamefont{Lukin}},
  \bibinfo{author}{\bibfnamefont{M.}~\bibnamefont{Rispoli}},
  \bibinfo{author}{\bibfnamefont{R.}~\bibnamefont{Schittko}},
  \bibinfo{author}{\bibfnamefont{P.~M.} \bibnamefont{Preiss}},
  \bibnamefont{and} \bibinfo{author}{\bibfnamefont{M.}~\bibnamefont{Greiner}},
  \bibinfo{journal}{Science} \textbf{\bibinfo{volume}{353}},
  \bibinfo{pages}{794} (\bibinfo{year}{2016}),
  \urlprefix\url{https://www.science.org/doi/abs/10.1126/science.aaf6725}.

\bibitem[{\citenamefont{Dymarsky et~al.}(2018)\citenamefont{Dymarsky, Lashkari,
  and Liu}}]{Dymarsky2018}
\bibinfo{author}{\bibfnamefont{A.}~\bibnamefont{Dymarsky}},
  \bibinfo{author}{\bibfnamefont{N.}~\bibnamefont{Lashkari}}, \bibnamefont{and}
  \bibinfo{author}{\bibfnamefont{H.}~\bibnamefont{Liu}},
  \bibinfo{journal}{Physical Review E} \textbf{\bibinfo{volume}{97}}
  (\bibinfo{year}{2018}),
  \urlprefix\url{https://doi.org/10.1103/PhysRevE.97.012140}.

\end{thebibliography}
\end{document}